# Blockchain and its Potential in Education

**Cristina Turcu[1], Cornel Turcu[1], Iuliana Chiuchişan[1]**

(1) Ştefan cel Mare University of Suceava
13, University Street, Suceava, RO-720229, ROMANIA
E-mail: cristina[at]eed.usv.ro

Abstract
*The proposed paper presents a literature review regarding the status of integrating the dynamic blockchain technology in the educational field. Blockchain is a relatively new technology and the same is its implementation in education. The emerging need in this area of research, which still is in its infancy, is justified by the possible use cases; some of these cases are in piloting phase, while others have already been adopted by educational institutions. This paper focuses on extending knowledge about blockchain and on identifying the benefits, risks and the associated challenges regarding the successful implementation of blockchain-based solutions in the field of education, fully in line with standards and guidelines for quality assurance.*

## 1  Introduction

The field of education is moving forward into the digital era. In fact, technology and education are a great combination that gains more popularity in recent years. Thus, education technology is becoming a global phenomenon, and according to (EdTechXGlobal, 2016), the market is projected to grow at 17.0% per annum, to $252 billion by 2020.

But we cannot talk about using technology without addressing the subject of security. Failure to comply with the appropriate security measures can result in the increased consumption of financial and human resources. Since the early adoption of technology in the field of education, researchers and practitioners have proposed different guidelines, approaches and policies to support the decision-making process regarding the security measures to be implemented. One solution that is gaining ground lately is based on blockchain technology, which offers powerful security capabilities. In recent years, there has been an increasing interest in this concept, justified by the exponential success of Bitcoin's cryptocurrency, launched by Satoshi Nakamoto (Nakamoto, S. 2009). Thus, the various possibilities of applying blockchain in different fields have caught the attention of researchers and practitioners. As the number of projects and applications based on this technology has increased, it is important that researchers and practitioners have access to the current state-of-the-art and –practice. In this work, a bibliographical research was carried out in order to understand the state of the art of applying blockchain technology in education. The aim of this paper is to provide a systematic classification and a synthesized overview of current practices. We have addressed the following research questions: (1) Is there evidence of the application of blockchain technology in education aimed at offering solutions to various issues? and (2) What are the reported outcomes? To achieve the proposed objective and to answer the research questions, we have performed a literature review study on the application of blockchain technology in the education field.

Blockchain, also known as Distributed Ledger Technology, revolutionizes the ways in which data is managed and transacted. In recent years, due to the many advantages, it began to be applied in various areas, all over the world, such as supply chain, transportation, logistics, public administration, etc. However, education, with some minor exceptions, is not currently considered a priority, even in countries with national blockchain initiatives. This is because, among other things, many stakeholders within education are currently unaware of the benefits and potential of blockchain technology.

But the adoption of technologies has a growing impact on the way skills are acquired and developed for the $21^{st}$ Century workforce. Thus, there is an increasing demand for skills update, which requires lifelong learning. But, sometimes, checking the authenticity of all diplomas and certificates that attest someone's skills and achievements proves difficult and time consuming. A system that allows an instant verification of the authenticity of these documents issued within the higher education area or other levels of education is paramount. Blockchain offers, besides high security, the capability to integrate data from disparate data sources, such as education records stored in databases of different educational providers.

The remainder of this paper is organized as follows. Section 2 presents the background and reviews the related work. The next section presents some blockchain-based solutions for education. Some of the benefits of adopting blockchain technology in education are showed in the Section 2.3. Next section discusses the limitations of blockchain technology in education, while section 3 states the future work directions. Finally, the conclusions are drawn in section 4.

## 2 Blockchain technology
### 2.1 State of the art

Blockchain technology is considered the key to solve scalability, privacy, and reliability problems in various domains (Malviya, 2016).

In short, a blockchain is a distributed digital ledger that enables the information recording and sharing by a community in which each member keeps his/her own copy of the information and must validate collectively any update (Piscini et al., 2016). The blockchain is a "trust-free, tamper-proof, auditable, and self-regulating system, with no human intervention required to execute computation" (Atzori, 2017). This encrypted database, that serves as an irreversible and incorruptible repository of information, "enables, for the first time, unrelated people to reach consensus on the occurrence of a particular transaction or event without the need for a controlling authority" (Wright & De Filippi, 2015).

Blockchain-based solutions are now being developed in many areas by big IT corporations such as IBM, Microsoft (Azure), Intel etc., but also by start-ups that are growing rapidly. According to (Grech & Camilleri, 2017), "the majority of EU Member States are likely to be experimenting with blockchain technologies. Some are working on national strategies, while others are conducting trials of specific application". Among these countries, there are Estonia, Netherlands, etc.

In this paper, we focus on the application of blockchain in education.

Blockcerts is an open standard for blockchain educational certificates (MIT Media Lab, 2016; Schmidt, 2015; Schmidt, 2017). In fact, this standard includes open-source libraries, tools, and mobile apps enabling a decentralized, standard-based, recipient-centric ecosystem, allowing the creation, issuing, holding, viewing and trustless verification of blockchain-based official records (Blockcerts, 2018). Currently, there are already commercial implementations of the blockcerts standard. An example of blockcerts application is MIT's Digital Diploma, that enables the issuing of secure digital diplomas, allowing students to share their verifiable and tamper-proof diploma digitally, free of charge and without involving an intermediate party (Durant and Trachy, 2017).

Due to the benefits of this technology, different blockchain-based platforms have recently been launched and their number is increasing. Therefore, we present some of the most popular platforms, focusing on the most suitable for developing applications in the educational field.

Some of the platforms for developing applications with a high impact lately are the following: Ethereum (Buterin et al., 2014), Hyperledger (Androulaki et al., 2018). Hyperledger is an open-source platform, on which various blockchain-based projects have been developed, including, for example, Hyperledger Fabric. The blockchain framework Hyperledger Fabric provides a distributed and scalable ledger on which other implementations, some of which commercial, such as the IBM Blockchain platform or Sony Global Education next-generation credentials platform (SGE, 2017) are based.

The blockchain is a relatively new technology and the application of blockchain in education is extremely new. Nevertheless, there already is a large number of blockchain-based proposals to improve some aspects of education. Some of these are presented in the next section.

**2.2 Blockchain-based solutions for education**

The blockchain technology is a transformative technology with an enormous potential. The scientific papers and reports published so far assert that, although education could greatly benefit from the functionality provided by blockchain, the adoption of block technology in education is in its infancy. The Joint Research Centre (JRC), the European Commission's science and knowledge service, published a report on Blockchain in Education (Grech & Camilleri, 2017) highlighting the potential and proposing some scenarios, including issuing certificates, verifying accreditation pathways, lifelong learning passports, intellectual property management, and data management. According to (Gräther et al., 2018), blockchain is highly suited for storing fingerprints of certificates or other educational items, due to the fact that each transaction is permanent recorded and verified.

Among the first blockchain-based systems, which moved from the prototype stage to commercial products, some refer to diplomas. These systems can be found, for example, at MIT, UT Austin, and the University of Nicosia (Cyprus), where digital diplomas are being issued to students.

In Europe, UK's Open University Knowledge Media Institute (KMI) is one of the pioneering universities to employ blockchain. KMI, in partnership with British Telecommunications (BT), has developed an Ethereum based platform for academic applications, named OpenBlockChain (Lemoie, 2017). For its experiments, the institute considers Microcredentials (badges) allocated for courses available on the Open Learn website and MOOCs (UK platform FutureLearn).

Open Badges are "verifiable, portable digital badges with embedded metadata about skills and achievements" (Mozilla Foundation, 2012) "that happen anytime, anywhere, and anyhow" (Lemoie, 2017). Badges can be degrees and certifications, but also microcredentials or any other type of credential.

Due to space limitations, we consider here only one use case related to education that can stand as proof for the important role blockchain can play in this field.

Usually, academic records of a person, such as degrees, diplomas, are separated data, stored in the databases of various providers of education and students or graduates do not have the authority to manage their own information. Moreover, no other unofficial person (e.g. an employer) has access to modify or even view these official records. This situation is common in most countries around the world, including Romania. But, in the context of heightened internationalization of education and work, with the increasing mobility of students and graduates, an easy access to the personal degree record is paramount. According to various studies, a significant percent of CVs contains false information regarding the person's academic track record. Thus, for example, (Risk Advisory Group, 2018), point out that "eighty percent contained one or more discrepancy - up by ten percent compared to our data last year. Discrepancies were most frequently related to educational qualifications and information about employment history". In the UK, there is a centralized service, named Higher Education Degree Datacheck (HEDD), that allows the verification of someone's degree. In Romania, the Ministry of National Education (Romanian acronym MEN) controls the management of the (Single) National Student Enrolment Registry (RMUR in Romanian), that is "a digital database that records all students in Romania from public and private universities, accredited or with a temporary license", in accordance with the Romanian Education Law no. 1/2011, art. 201, for all academic years and for all study cycles (RG, 2018). Keeping the record of all university diplomas issued in Romania, a strict control of the diplomas is ensured. But, so far, the access to this information is restricted, and students or graduates cannot view their own records from this central database. Moreover, a third-party, such as an employer, cannot access this database in order to verify the authenticity of a candidate's degrees.

Various studies identified the challenges regarding the recognition of degrees and diplomas issued by foreign education providers, taking into consideration the perils of degree counterfeiting, and made some recommendations for safeguarding against fraudulent documents (Trines S., 2017). One of these recommendations is a comprehensive and trusted system for recording, storing and retrieving educational information, such as, degrees, diplomas, education and training credentials, etc. Among other purposes, such a system would contribute to preventing fraud, by ensuring, for example, the management of both educational data and data access to third-parties (e.g. other universities, recruiters or employers), even from different countries.

But, "centralized data-storage and management systems are susceptible to hacking, intrusion, and breaches" (Efanov & Roschin, 2018). Instead, a distributed trust technology, ensuring scalability, privacy, and reliability, such as blockchain, could be considered. Thus, a blockchain-based model could be used for managing education data from various education service providers. This model will allow education providers to efficiently store and manage data, while ensuring data integrity and security. Also, data access with permission could be enabled. The universities and governments could be collective caretakers of the blockchain network. But only the universities should have access to create or update the data related to a student's degree record. Instead, the control of sharing this data should belong to the student/graduate, without requiring the permission of any official entities (universities, governments). This system should also address the quality assurance requirements.

The main benefits for stakeholders offered by such a system are depicted on the right side of figure 1.

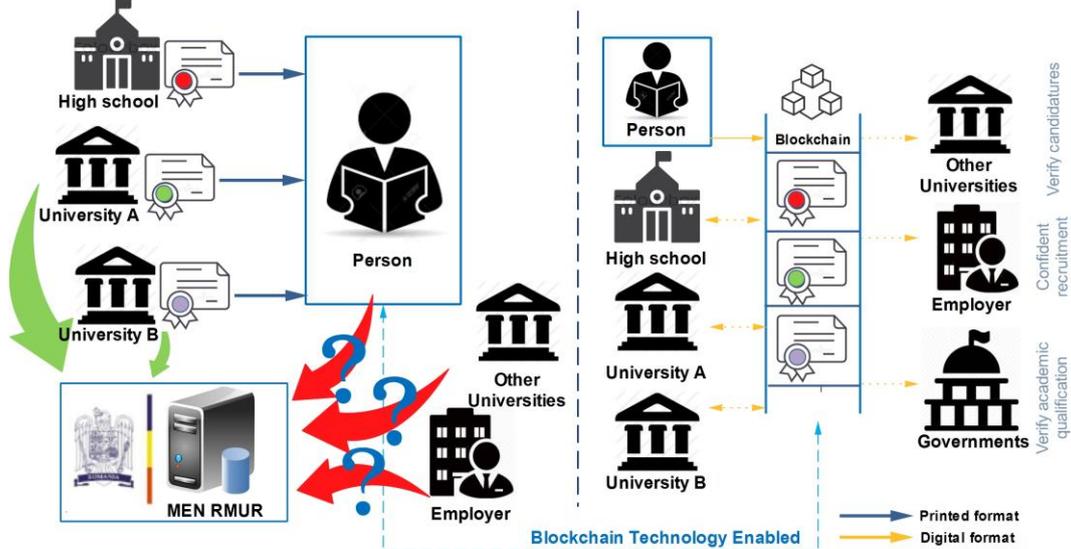

Figure 1. Romanian current state vs. blockchain-based solution

## 2. 3 Benefits of blockchain in education

Blockchain technology has limitless possibilities and could become an extensive part of education systems.

Some benefits of adopting blockchain technology in the field of education are the following (but are not limited to):
- decentralization: considering a P2P distributed architecture over a centralized one brings an improvement of the fault tolerance, by eliminating the central points of failures and bottlenecks (Veena et al., 2015);

- scalability: allows the elimination of situations where one or several entities control the storage and processing of information of a large number of people;
- reliability: the information can remain unchanged, immutable and distributed over time in blockchain. Any system participant can verify the authenticity of data and be certain it has not been tampered with (Reyna et al., 2018);
- security: information and communications can be secure if considered as transactions of the blockchain (Prisco, 2016), based on cryptographic protocols. Thus, for example, blockchain offers the potential to make degree records more secure.
- last but not least, the benefits that universities may have through the adoption of blockchain-based solutions include reducing administrative costs and bureaucracy.

**2.4 Limitations**

Some researchers and practitioners reported various issues that were encountered in the implementation of blockchain-based applications for education. Mainly, these revolve around GDPR compliance, data ownership and authenticity of data sources.

One of the limitations refers to the education providers joining the same blockchain network. One assumes that the identity of the education provider is verified by the system before joining the blockchain.

The European General Data Protection Regulation (GDPR), a robust and far-reaching privacy legislation, was created and adopted to impose the preservation of individual privacy and autonomy ("right to be forgotten" protections). Therefore, any solution must be GDPR compliant. But the blockchain removes "the need to trust a centralized authority in order to keep an accurate record of activity" and "it makes surveillance of activity extremely difficult. The blockchain represents the opportunity to not only fulfill but go beyond the promise of the European General Data Protection Regulation" (Smolenski, 2017). There are already some solutions, such as EvidenZ or BigchainDB, but any blockchain-based platform or application should carefully consider GDPR implications.

Latency is another limitation of blockchain, because the transaction can take too long.

Some assert that increasing storage capacity due to data redundancy (each node has a copy of the Blockchain) could also be an issue of blockchain adoption.

Moreover, many publications, including (Lin & Liao, 2017) reveal an important security issue of the blockchain-based system, the so-called "51% attack".

Also, we should emphasize the fact that the blockchain adoption in education makes it necessary to solve various legal issues.

In order to fully adopt blockchain technology in education ways in which these limitations could be alleviated or avoided altogether should be found.

**3 Future work**

Various reports and papers reveal the blockchain's potential to solve certain issues related to education, and also make various recommendations. But in order to fully exploit the blockchain in education, there are still technical and non-technical challenges that have to be addressed. Some of them are related to the blockchain technology itself (for example, overcoming current limitations, offering more powerful features and new capabilities), while others refer to education information systems. Thus, for example, some researchers consider the blockchain as a way to change in time the organization of the entire higher education system (McArthur David, 2018), in fact, revolutionizing higher education in order to create a global network for higher learning (Tapscott and Tapscott, 2017).

In Romania, considering the records from RMUR, MEN could support the development and implementation of a pilot project, based on the blockchain. In a first phase, this project could address PhD diplomas, that currently are all validated at the national level by the National Council for Titles, Diplomas, and Certificates.

## 4  Conclusion

In this paper, we examine outputs from the research and practitioner communities to identify how blockchain technology could be applied in the education field in order to potentially improve it.

Due to the benefits offered by this technology, various blockchain-based platforms have recently been launched and various applications have been developed in many diverse areas, including education, some of them being presented in this paper.

An important aspect is gaining access to important student or graduate-related data, which is often stored in the university database with exclusive access, or in the form of paper documents. Blockchain technology provides solutions to this issue, while allowing alignment with the emerging trend known as "open data".

Although the benefits of applying blockchain in the education field are many, this research topic is still in a preliminary stage and the adoption of standards and regulations is essential to expand its use. But it is worth noting that there are still other areas that have to be further investigated in order to fully adopt blockchain technology in education. Therefore the decision to apply the blockchain to this field must be carefully considered and taken with caution.